\def\ra{\rangle}
\def\la{\langle}
\def\be{\begin{equation}}
\def\ee{\end{equation}}
\def\ba{\begin{array}}
\def\ea{\end{array}}

\documentclass[aps,pra,preprint]{revtex4}

\begin{document}

\baselineskip=18pt \setcounter{page}{1} \centerline{\large\bf
Quantum Separability Criteria for Arbitrary}\medskip
\centerline{\large\bf Dimensional Multipartite States} \vspace{4ex}
\begin{center}
Ming Li$^{1,2}$, Jing Wang$^{1,3}$, Shao-Ming Fei$^{2,3}$ and
Xianqing Li-Jost$^{2}$

\vspace{2ex}

\begin{minipage}{5.5in}

{\small $~^{1}$ College of the Science, China University of
Petroleum, 266580 Qingdao}

{\small $~^{2}$ Max-Planck-Institute for Mathematics in the
Sciences, 04103 Leipzig}

{\small $~^{3}$ School of Mathematical Sciences, Capital Normal
University, 100048 Beijing}
\end{minipage}
\end{center}

\begin{center}
\begin{minipage}{5in}
\vspace{1ex} \centerline{\large Abstract} \vspace{1ex} We present
new separability criteria for both bipartite and multipartite
quantum states. These criteria include the criteria based on the
correlation matrix and its generalized form as special cases. We
show by detailed examples that our criteria are more powerful than
the positive partial transposition criterion, the realignment
criterion and the criteria based on the correlation matrices.

\vfill
\smallskip
\end{minipage}\end{center}
\bigskip

\section{Introduction}\label{sec1}

Quantum entanglement, as the remarkable nonlocal feature of quantum
mechanics, is recognized as a valuable resource in the rapidly
expanding field of quantum information science, with various
applications such as quantum computation \cite{nielsen, di}, quantum
teleportation \cite{teleportation}, dense coding \cite{dense},
quantum cryptographic schemes \cite{schemes}, quantum radar
\cite{malik}, entanglement swapping \cite{swapping} and remote state
preparation (RSP) \cite{RSP1,RSP2,RSP3,RSP4}. Quantum states without
entanglement are called separable states, which constitute a convex
subset of all the quantum states. Distinguishing quantum entangled states
from the separable ones is a basic and
longer standing problem in the theory of quantum entanglement. It has
attracted great interest in the last twenty years.

For mixed states we still have no general criterion. A strong
criterion, named PPT (partial positive transposition), to recognize
mixed entangled quantum state was proposed by Peres in 1996 in
\cite{ppt}. It says that for any bipartite separable quantum states
the density matrix must be semi-positive under partial transposition.
Afterwards, by using the method of positive maps the family Horodecki
\cite{ho96} showed that the Peres' criterion is also sufficient for
$2\times 2$ and $2\times3$ bipartite systems. For high-dimensional
states, the PPT criterion is only necessary. Horodecki \cite{ho232}
has constructed some classes of families of inseparable states with
positive partial transposes for $3\times 3$ and $2\times 4$ systems.
States of this kind are said to be bound entangled (BE). Another
powerful operational criterion for separability is the realignment
criterion \cite{ru02,ChenQIC03}. It demonstrates a remarkable
ability in detecting the entanglement of many bound entangled states and even genuinely
tripartite entanglement \cite{3h0206008}. Considerable efforts have
been made in proposing stronger variants and multipartite
generalizations for this criterion
\cite{generalizerealignment,chenkai}. It was shown that PPT
criterion and realignment criterion are equivalent to the
permutations of the density matrix's indices \cite{3h0206008}.

Recently, some more elegant results for the separability problem have
been derived. In \cite{hofmann,guhne1,117903}, a
separability criteria based on the local uncertainty relations (LUR)
was obtained. The authors show that for any separable state
$\rho\in{\cal{H}}_A\otimes{\cal{H}}_B$,
$$
1-\sum\limits_{k}\la
G_{k}^{A}\otimes G_{k}^{B}\ra-\frac{1}{2}\la G_{k}^{A}\otimes I -
I\otimes G_{k}^{B}\ra^{2}\geq 0,
$$
where $G_{k}^{A}$ or $G_{k}^{B}$ are arbitary local orthogonal and
normalized operators (LOOs) in ${\cal{H}}_A\otimes{\cal{H}}_B$. This
criterion is strictly stronger than the realignment criterion. Thus
more bound entangled quantum states can be recognized by the LUR
criterion. The criterion is optimized in \cite{optloo} by choosing
the optimal LOOs. The covariance matrix of a quantum state is also
used to study separability in \cite{guhne}. It has been pointed out
in \cite{guhnecov2} that the LUR criterion, including the optimized
one, can be derived from the covariance matrix criterion. In
\cite{vicente1} the author has given a criterion based on the
correlation matrix of a state. The correlation matrix (CM) criterion
is then shown to be independent of PPT and realignment criterion in
\cite{vicente2}, i.e. there exist quantum states that can be
recognized by the correlation criterion while the PPT, realignment
criterion and the covariance matrix criterion fail. In
\cite{vicente3}, by defining matricizations of the correlation
tensors, the authors introduced a general framework for
detecting genuine multipartite entanglement and non-fully
separability in multipartite quantum systems.

In this paper, we present a generalized form of the correlation
matrix criterion for bipartite quantum systems
\cite{vicente1,vicente2} and for multipartite quantum systems
\cite{hassan}. Our new criterion includes the criterion based on the
correlation matrix as a special case and is more powerful than the
later for detecting entanglement, as shown by detailed examples.
Thus our criterion will be more efficient than the Positive partial
transposition criterion, the realignment criterion and the
covariance matrix criterion for some quantum states.

\section{Separability criterion for bipartite quantum states}\label{sec2}

Let $H_A^{d_1}$ and $H_B^{d_2}$ be two vector spaces with dimensions
${d_1}$ and ${d_2}$ respectively. By using the generators of
$SU(d)$, $\lambda_i$, $i=1,2,...,d^2-1$, any quantum state $\rho\in
H_A^{d_1}\otimes H_B^{d_2}$ can be writing as: \be\label{1}
\rho=\frac{1}{d_1d_2}I\otimes
I+\sum_{k=1}^{d_1^2-1}r_k\lambda_k\otimes I+\sum_{l=1}^{d_2^2-1}s_l
I\otimes\lambda_l+\sum_{k=1}^{d_1^2-1}\sum_{l=1}^{d_2^2-1}t_{kl}\lambda_k\otimes
\lambda_l, \ee where $r_k=\frac{1}{2d_2}Tr(\rho\lambda_k\otimes I),
s_l=\frac{1}{2d_1}Tr(\rho I\otimes\lambda_l)$ and
$t_{kl}=\frac{1}{4}Tr(\rho\lambda_k\otimes\lambda_l)$. We denote $T$
the matrix with entries $t_{kl}$ and define
\be\label{dt}\tilde{T}=\left(%
    \begin{array}{ccccc}
      \frac{1}{d_1d_2} & s_1 & s_2 & \cdots & s_{d_2^2-1} \\
      r_1 & t_{11} & t_{12} & \cdots & t_{1(d_2^2-1)} \\
      r_2 & t_{21} & t_{22} & \cdots & t_{2(d_2^2-1)} \\
      \cdots\\
      r_{d_1^2-1} & t_{(d_1^2-1)1} & t_{(d_1^2-1)2} & \cdots & t_{(d_1^2-1)(d_2^2-1)}
    \end{array}%
    \right).
   \ee

{\bf{Theorem 1:}} If $\rho\in H_A^{d_1}\otimes H_B^{d_2}$ is
separable, then for any $d_1^2\otimes d_2^2$ matrix $M$ and
$(d_1^2-1)\otimes (d_2^2-1)$ matrix $N$ with real entries $m_{ij}$
and $n_{ij}$ respectively,
\begin{eqnarray}\label{3e}|\sum_{kl}m_{kl}\widetilde{T}_{kl}|&\leq&
\frac{\sqrt{(d_1^2-d_1+2)(d_2^2-d_2+2)}}{2d_1d_2}\sigma_{max}(M)\\
|\sum_{kl}n_{kl}t_{kl}|&\leq&
\sqrt{\frac{(d_1-1)(d_2-1)}{4d_1d_2}}\sigma_{max}(N),
\end{eqnarray}
where $\sigma_{max}(M)$ and $\sigma_{max}(N)$ are the maximal
singular values of $M$ and $N$ respectively.

{\bf{Proof:}}  A separable quantum state $\rho$ can be expressed as:
\be
\rho=\sum_ip_i|\psi_i\ra\la\psi_i|\otimes|\phi_i\ra\la\phi_i|.
\ee
By writing the pure states $|\psi_i\ra$ and $|\phi_i\ra$ in
their Bloch forms, we have that
\begin{eqnarray}\label{2}
\rho&=&\sum_ip_i|\psi_i\ra\la\psi_i|\otimes|\phi_i\ra\la\phi_i|\nonumber\\
&=&\sum_ip_i(\frac{1}{d_1}I+\sum_kx_{ik}\lambda_k)\otimes(\frac{1}{d_2}I+\sum_ly_{il}\lambda_l)\nonumber\\
&=&\frac{1}{d_1d_2}I\otimes
I+\frac{1}{d_2}\sum_ip_i\sum_kx_{ik}\lambda_k\otimes
I+\frac{1}{d_1}\sum_ip_i\sum_ly_{il}I\otimes\lambda_k\nonumber\\
&&+\sum_ip_i\sum_{kl}x_{ik}y_{il}\lambda_k\otimes\lambda_l.
\end{eqnarray}
Comparing (\ref{1}) with (\ref{2}), we have \be\label{3}
r_k=\frac{1}{d_2}\sum_ip_ix_{ik},~~~
s_l=\frac{1}{d_1}\sum_ip_iy_{il},~~~
t_{kl}=\sum_ip_i\sum_{kl}x_{ik}y_{il}. \ee

Define $\vec{\tilde{x}}_i=(\frac{1}{d_1},x_{i1},\cdots,
x_{i(d_1^2-1)})^t$ and
$\vec{\tilde{y}}_i=(\frac{1}{d_2},y_{i1},\cdots, y_{i(d_2^2-1)})^t,$
where $t$ stands for the transposition. Since $|\psi_i\ra\in
H_A^{d_1}$ and $|\phi_i\ra\in H_B^{d_2}$ are all pure states, one
has \be\label{pn}
Tr(|\psi_i\ra\la\psi_i)^2=Tr(\frac{1}{d_1}I+\sum_kx_{ik}\lambda_k)^2=\frac{1}{d_1}+2\sum_kx_{ik}^2=1,
\ee i.e.
$||\vec{x}_i||=\sqrt{\sum_kx_{ik}^2}=\sqrt{\frac{d_1-1}{2d_1}}$.
Hence $||\vec{\tilde{x}}_i||=\sqrt{\frac{d_1^2-d_1+2}{2d_1^2}}$.
Similarly we have
$||\vec{\tilde{y}}_i||=\sqrt{\frac{d_2^2-d_2+2}{2d_2^2}}$. Therefore
for any real matrices $M$ and $N$, one obtains that \be
|\sum_{kl}m_{kl}\widetilde{T}_{kl}|=|\sum_{ikl}p_im_{kl}\tilde{x}_{ik}\tilde{y}_{il}|
\leq \sum_{i}p_i|\la\vec{\tilde{x}}_{i},M\vec{\tilde{y}}_{i}\ra|\leq
\frac{\sqrt{(d_1^2-d_1+2)(d_2^2-d_2+2)}}{2d_1d_2}\sigma_{max}(M)\nonumber;
\ee \be\label{lhvo}
|\sum_{kl}n_{kl}t_{kl}|=|\sum_{ikl}p_in_{kl}x_{ik}y_{il}| \leq
\sum_{i}p_i|\la\vec{x}_{i},N\vec{y}_{i}\ra|\leq
\sqrt{\frac{(d_1-1)(d_2-1)}{4d_1d_2}}\sigma_{max}(N). \nonumber\ee
\hfill\rule{1ex}{1ex}

The correlation matrix criterion in \cite{vicente1} illustrates that
if quantum state $\rho$ is separable, then the Key-Fan norm $||T||_{KF}\leq
\sqrt{\frac{(d_1-1)(d_2-1)}{4d_1d_2}}$. In the following we show the power of Theorem 1
in detecting entanglement by two corollaries.

{\bf{Corollary 1:}} The criterion based on the correlation matrix is
included in Theorem 1.

{\bf{Proof:}} Let $T=U\Sigma V^{\dag}$ be the singular value
decomposition of $T$. Since $T$ is a real matrix, one can always
choose $U$ and $V$ to be orthogonal matrices. Without loss of
generality, we assume that $d_1\leq d_2$. Set $N=(V\Delta
U^{\dag})^t$, where $\Delta$ is a block matrix of the form $\left(%
    \begin{array}{cc}
      I & 0\\
      \end{array}%
    \right)^t$, $I$ is the $(d_1^2-1)\times (d_1^2-1)$ identity matrix,
$0$ stands for a $(d_2^2-d^2_1)\times (d_2^2-d_1^2)$ zero
    matrix. The singular values of $N$ must be either $1$ or $0$. One
    obtains
\begin{eqnarray*}
||T||_{KF}&=&|Tr(U\Sigma V^{\dag}V\Delta
U^{\dag})|=|Tr(TN^t)|=|\sum_{kl}n_{kl}t_{kl}|\\
&\leq&\sqrt{\frac{(d_1-1)(d_2-1)}{4d_1d_2}}\sigma_{max}(N)=\sqrt{\frac{(d_1-1)(d_2-1)}{4d_1d_2}}.
\end{eqnarray*}
This means that one can get the correlation matrix criterion from
Theorem 1. \hfill \rule{1ex}{1ex}

{\bf{Corollary 2:}} If a bipartite quantum state $\rho\in
H_A^{d_1}\otimes H_B^{d_2}$ is separable, then the following
inequality must hold: \be||\tilde{T}||_{KF}\leq
\frac{\sqrt{(d_1^2-d_1+2)(d_2^2-d_2+2)}}{2d_1d_2},\ee where
$||\Omega||_{KF}=Tr\sqrt{\Omega\Omega^{\dag}}$ stands for the trace
norm of $\Omega$.

{\bf{Proof:}} Assume $d_1\leq d_2$. Let
$\tilde{T}=X\Sigma Y^{\dag}$ be the singular value decomposition of
$\tilde{T}$, with $X$ and $Y$ the corresponding orthogonal matrices.
Set $M=(Y\Gamma X^{\dag})^t$, where $\Gamma=\left(%
    \begin{array}{cc}
      I & 0\\
      \end{array}%
    \right)^t$, $I$ and $0$ are the $d_1^2\times d_1^2$ identity matrix
    and the $(d_2^2-d^2_1)\times (d_2^2-d_1^2)$ zero
    matrix, respectively. The singular values
of $M$ are either $1$ or $0$. Then we obtain that
\begin{eqnarray*}
||\tilde{T}||_{KF}&=&|Tr(X\Sigma Y^{\dag}Y\Gamma
X^{\dag})|=|Tr(\tilde{T}M^t)|=|\sum_{kl}m_{kl}\tilde{T}_{kl}|\\&\leq&
\frac{\sqrt{(d_1^2-d_1+2)(d_2^2-d_2+2)}}{2d_1d_2}\sigma_{max}(M)=\frac{\sqrt{(d_1^2-d_1+2)(d_2^2-d_2+2)}}{2d_1d_2},
\end{eqnarray*}
which ends the proof of the corollary.  \hfill \rule{1ex}{1ex}

Corollary 1 shows that Theorem 1 is not weaker than the
correlation matrix criterion in detecting entanglement for quantum
states in $H_A^{d_1}\otimes H_B^{d_2}$. In fact, by the following
example we can show that Theorem 1 is strictly stronger than
the correlation matrix criterion, the realignment criterion and
the PPT criterion.

{\bf{Example:}} A $3\times 3$ PPT entangled state is given in \cite{bennett}:
\be
\rho=\frac{1}{4}(I_9-\sum_{i=0}^4|\psi_i\ra\la\psi_i|), \ee where
$|\psi_0\ra=|0\ra(|0\ra-|1\ra)/\sqrt{2}$,
$|\psi_1\ra=(|0\ra-|1\ra)|2\ra/\sqrt{2}$,
$|\psi_2\ra=|2\ra(|1\ra-|2\ra)/\sqrt{2}$,
$|\psi_3\ra=(|1\ra-|2\ra)|0\ra/\sqrt{2}$ and
$|\psi_4\ra=(|0\ra+|1\ra+|2\ra)(|0\ra+|1\ra+|2\ra)/3$. The state is
shown to violate the correlation matrix criterion. Let us mix $\rho$
with white noise: \be \sigma(x)=x\rho+\frac{1-x}{9}I_9. \ee The
correlation matrix criterion detects the entanglement for
$0.9493<x\leq 1$. If we choose the matrix $M$ in theorem 1 to be
\begin{eqnarray*}
\left(%
    \begin{array}{ccccccccc}
    0.8134 & 0.1905& -0.11& 0.18& -0.4067& 0.1798& 0& 0&
  0\\
  0.1905& 0.3849& -0.243& -0.806& 0.2608& -0.0989&
  0& 0& 0\\
  -0.11& -0.243& 0.1043& -0.3511& -0.1506&
  0.8736& 0& 0& 0\\
  0.1798& -0.0989&
  0.8736& -0.3258& -0.1634& -0.2898& 0& 0& 0\\
  -0.4067&
  0.2608& -0.1506& -0.1634& -0.867& -0.1634& 0& 0&
  0\\
  0.1798& -0.806& -0.3511& -0.2898& -0.1634& -0.3258&
   0& 0& 0\\
   0& 0& 0& 0& 0& 0& 0.964& 0& 0\\
   0& 0& 0& 0& 0& 0& 0&
  0.964& 0\\
  0& 0& 0& 0& 0& 0& 0& 0& 0.964\\
     \end{array}%
    \right),
\end{eqnarray*}
which has the maximal singular value 1.036. From (\ref{3e}) the state $\sigma(x)$
is entangled for $0.94<x\leq 1$. Furthermore, by corollary 2 one can show that
$\sigma(x)$ is entangled for  $0.89254<x\leq 1$. Here one finds that our criterion is much better
than the correlation matrix criterion.

\section{Separability criterion for multipartite quantum states}\label{sec2}

In this section we consider the separability problem for N-partite
quantum systems $H_1\otimes H_2\otimes\cdots\otimes H_N$ with
$dim\,H_i=d_i$, $i=1,2,\cdots,N$.

Let $\lambda_{\alpha_{k}}^{\{\mu_{k}\}}=I_{d_{1}}\otimes
I_{d_{2}}\otimes\cdots\otimes \lambda_{\alpha_{k}}\otimes
I_{d_{\mu_{k}+1}}\otimes\cdots\otimes I_{d_{N}}$ with
$\lambda_{\alpha_{k}}$, the generators of  $SU(d_i)$, appearing at
the $\mu_k$th position and
\begin{eqnarray*}
{\mathcal{T}}_{\alpha_{1}\alpha_{2}\cdots\alpha_{M}}
^{\{\mu_{1}\mu_{2}\cdots\mu_{M}\}}=\frac{\prod_{i=1}^{M}
d_{\mu_{i}}}{2^{M}\Pi_{i=1}^{N}d_{i}}{\rm
Tr}[\rho\lambda_{\alpha_{1}}
^{\{\mu_{1}\}}\lambda_{\alpha_{2}}^{\{\mu_{2}\}}\cdots\lambda_{\alpha_{M}}^{\{\mu_{M}\}}],
\end{eqnarray*}
which can be viewed as the entries of the tensors
${\mathcal{T}}^{\{\mu_{1}\mu_{2}\cdots\mu_{M}\}}$.

For $\alpha_{M}=\cdots=\alpha_N=0$ with $1\leq M\leq N$, we define
that$\tilde{{\mathcal{T}}}_{\alpha_1\alpha_2\cdots\alpha_N}={\mathcal{T}}^{\mu_1\cdots\mu_M}_{\alpha_1\cdots\alpha_M},
$ and for $\alpha_{1}=\cdots=\alpha_N=0$, define that
$\tilde{{\mathcal{T}}}_{\alpha_1\cdots\alpha_N}=\frac{1}{\Pi_{k=1}^Nd_k}$.
Hence we have a tensor $\tilde{{\mathcal{T}}}$ with elements
$\{\tilde{{\mathcal{T}}}_{\alpha_1\cdots\alpha_N},\,\alpha_k=0,1,\cdots,d_k^2-1\}$.

If we set $\lambda_0^{\{k\}}=I_{d_k}$ for any $1\leq k \leq N$, then
any multipartite state $\rho\in H_1\otimes H_2\otimes\cdots\otimes
H_N$ can be generally expressed by the tensor
$\tilde{{\mathcal{T}}}$ as \cite{hassan}, \be\label{OS}
\rho=\sum\limits_{\alpha_{1}\alpha_{2}\cdots\alpha_{N}}
\tilde{{\mathcal{T}}}_{\alpha_{1}\alpha_{2}\cdots\alpha_{N}}\lambda_{\alpha_{1}}
^{\{1\}}\lambda_{\alpha_{2}}^{\{2\}}\cdots\lambda_{\alpha_{N}}^{\{N\}},
\ee where the summation is taken for all
${\alpha_{k}}=0,1,\cdots,d_k^2-1$.

To obtain the criterion for N-partite quantum systems, we adopt the
definition of the $n$th matrix unfolding ${\mathcal{T}}^n$ of a
tensor ${\mathcal{T}}$, which is a matrix with $i_n$ to be the row
index and the rest subscripts of ${\mathcal{T}}$ to be column
indices(detailed description can be found in Refs.
\cite{siam,hassan}). The Ky Fan norm of the tensor ${\mathcal{T}}$
over N matrix unfoldings is defined as \be
||{\mathcal{T}}||_{KF}=\max\{||{\mathcal{T}}_n||_{KF}\},~~~
n=1,2,\cdots,N. \ee

{\bf{Theorem 2:}} If a quantum state $\rho\in H_1\otimes
H_2\otimes\cdots\otimes H_N$ is fully separable, then for any
tensors $M$ and $W$ with real entries $m_{i_1i_2\cdots i_N}$,
$i_k=1,2,\cdots,d_k^2-1$, and $w_{j_1j_2\cdots j_N}$,
$j_l=1,2,\cdots,d_k^2$, we have:
\begin{eqnarray}
|\sum_{i_1i_2\cdots i_N}m_{i_1i_2\cdots
i_N}{\mathcal{T}}_{i_1i_2\cdots i_N}|\leq
\Pi_{k=1}^N\sqrt{\frac{d_k-1}{2d_k}}\sigma_{max}(M),
\end{eqnarray}
\begin{eqnarray}\label{mu2}
|\sum_{j_1j_2\cdots j_N}w_{j_1j_2\cdots
j_N}\tilde{{\mathcal{T}}}_{i_1i_2\cdots i_N}|\leq
\Pi_{k=1}^N\sqrt{\frac{d_k^2-d_k+2}{2d_k^2}}\sigma_{max}(W),
\end{eqnarray}
where $\sigma_{max}(M)$ and $\sigma_{max}(W)$ stand for the maximal
eigenvalue of the matrix unfolding $M_n$ and $W_n$. The maximum is
taken over all kinds of mode n matricization.

{\bf{Proof:}} Assume that $\rho\in H_1\otimes
H_2\otimes\cdots\otimes H_N$ is fully separable, one can always find
the following decomposition:
\be\label{5}
\rho=\sum_ip_i|\psi^1_i\ra\la\psi^1_i|\otimes
|\psi^2_i\ra\la\psi^2_i|\otimes\cdots\otimes
|\psi^N_i\ra\la\psi^N_i|,
\ee
where $|\psi^m_i\ra\la\psi^m_i|$ are
density matrices of pure states in $H_m$. Using
the Bloch representation of density matrix, we have that
\be\label{4}
|\psi^m_i\ra\la\psi^m_i|=\frac{1}{d_m}I+\sum_{\alpha_m}
x^m_{i\alpha_m}\lambda_{\alpha_m},
\ee
where $x^m_{i\alpha_m}={Tr(|\psi^m_i\ra\la\psi^m_i|\lambda_{\alpha_m})}/{2}$.
By (\ref{pn}) one has that
$||\vec{x}_i^m||=\sqrt{\frac{d_m-1}{2d_m}}$. Denote
$\vec{\tilde{x}}_i^m=(\frac{1}{d_m},x_{i1}^m,\cdots,
x_{i(d_1^2-1)}^m)^t$. We obtain that
$||\vec{\tilde{x}}_i^m||=\sqrt{\frac{d_m^2-d_m+2}{2d_m^2}}$.
Substituting (\ref{4}) into (\ref{5}) one has that:
\begin{eqnarray}\label{3x}
\rho&=&\frac{1}{\Pi_{k=1}^Nd_k}\otimes_{k=1}^N
I_k+\sum_{\mu_1\alpha_1}\frac{d_{\mu_1}}{\Pi_{k=1}^N}\sum_ip_ix^{\mu_1}_{i\alpha_1}\lambda^{\mu_1}_{\alpha_1}
+\sum_{\mu_1\mu_2\alpha_1\alpha_2}\frac{d_{\mu_1}d_{\mu_2}}{\Pi_{k=1}^N}\sum_ip_ix^{\mu_1}_{i\alpha_1}x^{\mu_2}_{i\alpha_2}
\lambda^{\mu_1}_{\alpha_1}\lambda^{\mu_2}_{\alpha_2}\nonumber\\
&&+\cdots+\sum_{\mu_1\cdots\mu_M,\alpha_1\cdots\alpha_M}\frac{\Pi_{k=1}^Md_{\mu_k}}{\Pi_{k=1}^N}\sum_ip_ix^{\mu_1}_{i\alpha_1}\cdots
x^{\mu_M}_{i\alpha_M}
\lambda^{\mu_1}_{\alpha_1}\cdots\lambda^{\mu_M}_{\alpha_M}\nonumber\\
&&+\sum_{\alpha_1\cdots\alpha_N}\sum_ip_ix^{1}_{i\alpha_1}\cdots
x^{N}_{i\alpha_N} \lambda^{1}_{\alpha_1}\cdots\lambda^{N}_{\alpha_N}.
\end{eqnarray}
Comparing (\ref{OS}) and (\ref{3x}), one gets
\be{\mathcal{T}}_{\alpha_{1}\alpha_{2}\cdots\alpha_{M}}^{\{\mu_{1}\mu_{2}\cdots\mu_{M}\}}
=\frac{\Pi_{k=1}^Md_{\mu_k}}{\Pi_{k=1}^N}\sum_ip_ix^{\mu_1}_{i\alpha_1}\cdots
x^{\mu_M}_{i\alpha_M}.\ee According to the definitions of
$\vec{x}_i^m, \vec{\tilde{x}}_i^m$ and
${\mathcal{T}}_{\alpha_1\alpha_2\cdots\alpha_N},
\tilde{{\mathcal{T}}}_{\alpha_1\alpha_2\cdots\alpha_N}$, we have
that
\begin{eqnarray}{\mathcal{T}}_{\alpha_{1}\alpha_{2}\cdots\alpha_{N}}
=\sum_ip_ix^1_{i\alpha_1}\cdots
x^N_{i\alpha_N}=\sum_ip_i\vec{x}^1_i\circ\vec{x}^2_i\circ\cdots\circ\vec{x}^N_i\\
\tilde{{\mathcal{T}}}_{\alpha_{1}\alpha_{2}\cdots\alpha_{N}}
=\sum_ip_i\tilde{x}^1_{i\alpha_1}\cdots
\tilde{x}^N_{i\alpha_N}=\sum_ip_i\vec{\tilde{x}}^1_i\circ\vec{\tilde{x}}^2_i\circ\cdots\circ\vec{\tilde{x}}^N_i,\end{eqnarray}
where $\circ$ stands for the out product.

Let $M_n$ be mode n matricization of $M$. Then for any tensor $M$ we
have that
$$
\sum_{i_1i_2\cdots i_N}m_{i_1i_2\cdots
i_N}{\mathcal{T}}_{i_1i_2\cdots i_N}=\sum_ip_i\la
\vec{x}_i^n,M_n(\vec{x}^1_i\circ\cdots\circ\vec{x}^{\hat{n}}_i\circ\cdots\circ\vec{x}^N_i)^t\ra\leq
\Pi_{k=1}^N\sqrt{\frac{d_k-1}{2d_k}}\sigma_{max}(M).
$$

Inequality (\ref{mu2}) can be derived similarly. \hfill
\rule{1ex}{1ex}

In \cite{hassan}, the authors have derived a generalized form of the
correlation matrix criterion which says that if a quantum state
$\rho\in H_1\otimes H_2\otimes\cdots\otimes H_N$ is fully separable,
then
\begin{eqnarray}\label{gcmc}||{\mathcal{T}}||_{KF}&=&||{\mathcal{T}}_{n}||_{KF}
\leq\Pi_{k=1}^N\sqrt{\frac{d_k-1}{2d_k}}.\end{eqnarray}

Here we show that one can obtain the generalized correlation matrix
criterion from Theorem 2.

{\bf{Corollary 3:}} Inequality (\ref{gcmc}) is included in theorem
2. Moreover, if quantum state $\rho\in H_1\otimes
H_2\otimes\cdots\otimes H_N$ is fully separable, then the following
inequality holds:
\begin{eqnarray}\label{ggcmc}||\tilde{{\mathcal{T}}}||_{KF}&=&||\tilde{{\mathcal{T}}}_{n}||_{KF}
\leq\Pi_{k=1}^N\sqrt{\frac{d_k^2-d_k+2}{2d_k^2}}.\end{eqnarray}

{\bf{Proof:}} Assume that the $n$th unfold ${\mathcal{T}}_n$ is
just the one to attain the $||{\mathcal{T}}||_{KF}$. One immediately
derives a singular value decomposition of ${\mathcal{T}}_n$,
${\mathcal{T}}_n=V_n\Sigma_n U_n^{\dag}$ for some orthogonal
matrices $V_n$ and $U_n$. Let $M$ be the tensor with the $n$th
matrix unfolding $M_n=V_n\Pi_n U_n^{\dag}$, where $\Pi_n=\left(%
    \begin{array}{cc}
      I & 0\\
      \end{array}%
    \right)$, $I$ is the $(d_n^2-1)\times (d_n^2-1)$ identity matrix
    and $0$ is the zero matrix with order such that $\Pi_n$ is a $(d_n^2-1)\times \frac{\prod_{k=1}^N(d_k^2-1)}{(d_n^2-1)}$ matrix.
Since both $V_n$ and $U_n$ are orthogonal matrices, the maximal singular value must be $1$. From Theorem 2 we have
\begin{eqnarray*}
&&|\sum_{i_1i_2\cdots i_N}m_{i_1i_2\cdots
i_N}{\mathcal{T}}_{i_1i_2\cdots i_N}|=
Tr(M_nT_n^{\dag})=Tr(V_n\Pi_nU_n^{\dag}U_n\Sigma_nV_n^{\dag})\\
&=&Tr(\Sigma_n)=||{\mathcal{T}}||_{KF}\leq
\Pi_{k=1}^N\sqrt{\frac{d_k-1}{2d_k}},
\end{eqnarray*}
which leads to the inequality (\ref{gcmc}). Inequality (\ref{ggcmc})
can be proved similarly.\hfill \rule{1ex}{1ex}

Corollary 3 can detect some PPT entangled quantum states in
multipartite quantum systems, such as the three-qutrit bound
entangled states $\rho_c\otimes |\psi\ra\la\psi|$ condidered by L.
Clarisse and P. Wocjan \cite{wocjan}, where
\begin{eqnarray*}
\rho_c=\frac{1}{12}\left(%
    \begin{array}{ccccccccc}
   1&0&1&0&0&0&1&0&0\\
  0&1&0&0&0&-1&0&-1&0\\
  1&0&2&0&-1&0&0&0&0\\
  0&0&0&1&0&-1&0&1&0\\
  0& 0& -1& 0& 1& 0& 1& 0& 0\\
  0 &-1& 0 &-1& 0& 2& 0& 0& 0\\
  1& 0& 0& 0& 1& 0& 2& 0& 0\\
  0&-1& 0& 1& 0& 0& 0& 2& 0\\
  0& 0& 0& 0& 0& 0& 0& 0& 0\\
     \end{array}%
    \right)
\end{eqnarray*}
is the chess-board state and $|\psi\ra$ is an uncorrelated ancilla.
If we mix $\rho_c\otimes |\psi\ra\la\psi|$ with white noise and
define $\sigma=p\rho_c\otimes |\psi\ra\la\psi|+\frac{1-p}{27}I$, the
entanglement is detected for $0.83265<p\leq1$ by corollary 3.

\section{Conclusions and Remarks}\label{sec5}

It is a basic and fundamental question to distinguish separable
quantum states from entangled ones. Although the quantum
separability problem has been shown to be NP-hard, it is possible to
derive some necessary criteria of separability. We have derived
separability criteria of quantum states for both bipartite and
multipartite quantum ones. The criteria are shown to be more
efficient in detecting quantum entanglement of some quantum states
than the (generalized) criterion based on the correlation matrix,
the PPT criterion, the realignment criterion, and the covariance
matrix criterion. Similar to the case of previous separability
criteria, our criteria can also be used to derive lower bounds for
concurrence.

\bigskip
\noindent{\bf Acknowledgments}\, \, This work is supported by the
NSFC 11105226, 11275131; the Fundamental Research Funds for the
Central Universities No.12CX04079A, No.24720122013; Research Award
Fund for outstanding young scientists of Shandong Province
No.BS2012DX045.

\end{document}